\documentstyle[12pt]{ioplppt}          
\begin{document}
\title{Harmonic inversion as a general method for periodic orbit quantization}

\author{J\"org Main,$^1$ Vladimir A. Mandelshtam,$^2$ G\"unter Wunner,$^3$
and Howard S. Taylor$^4$}

\address{$^1$ Theoretische Physik I, Ruhr-Universit\"at Bochum, 
         D-44780 Bochum, Germany}
\address{$^2$ University California Irvine, Department of Chemistry, 
         Los Angeles, CA 92612}
\address{$^3$ Inst. f. Theor. Physik und Synergetik, Universit\"at Stuttgart,
         D-70550 Stuttgart, Germany}
\address{$^4$ University of Southern California, Department of Chemistry, 
         Los Angeles, CA 90089}
\def\today{}

\begin{abstract}
In semiclassical theories for chaotic systems such as Gutzwiller's periodic
orbit theory the energy eigenvalues and resonances are obtained as poles
of a non-convergent series $g(w)=\sum_n A_n\exp(is_nw)$.
We present a general method for the analytic continuation of such 
a non-convergent series by harmonic inversion of the ``time'' signal, which 
is the Fourier transform of $g(w)$.
We demonstrate the general applicability and accuracy of the method on two
different systems with completely different properties: the Riemann zeta 
function and the three disk scattering system.
The Riemann zeta function serves as a mathematical model for a bound system.
We demonstrate that the method of harmonic inversion by filter-diagonalization
yields several thousand zeros of the zeta function to about 12 digit 
precision as eigenvalues of small matrices.
However, the method is not restricted to bound and ergodic systems, and does 
not require the knowledge of the mean staircase function, i.e., the Weyl term 
in dynamical systems, which is a prerequisite in many semiclassical 
quantization conditions.
It can therefore be applied to open systems as well.
We demonstrate this on the three disk scattering system, as a physical example.
The general applicability of the method is emphasized by the fact that one 
does not have to resort a symbolic dynamics, which is, in turn, the basic 
requirement for the application of cycle expansion techniques.

\end{abstract}


\maketitle

\section{Introduction}
Since the development of {\em periodic orbit theory} by Gutzwiller 
\cite{Gut67,Gut90} it has become a fundamental question as to how individual
semiclassical eigenenergies and resonances can be obtained from periodic
orbit quantization for classically chaotic systems.
A major problem is the exponential proliferation of the number of periodic 
orbits with increasing period, resulting in a divergence of 
Gutzwiller's trace formula at real energies and below the real axis, where 
the poles of the Green's function are located.
The periodic orbit sum is a Dirichlet series
\begin{equation}
 g(w) = \sum_n A_n e^{i s_n w} \; ,
\label{g}
\end{equation}
where the parameters $A_n$ and $s_n$ are the amplitudes and periods (actions)
of the periodic orbit contributions.
In most applications Eq.\ \ref{g} is absolutely convergent only in the region 
${\rm Im}~w > c > 0$ with $c$ the entropy barrier of the system, while the 
poles of $g(w)$, i.e., the bound states and resonances, are located on and 
below the real axis, ${\rm Im}~w \le 0$.
Thus, to extract individual eigenstates, the semiclassical trace formula 
(\ref{g}) has to be analytically continued to the region of the quantum poles.

Up to now no general procedure is known for the analytic continuation of a
non-convergent Dirichlet series of the type of Eq.\ \ref{g}.
All existing techniques are restricted to special situations.
For bound and ergodic systems the semiclassical eigenenergies can be 
extracted with the help of a functional equation and the mean staircase 
function (Weyl term), resulting in a Riemann-Siegel look-alike formula 
\cite{Ber86,Ber90,Kea92,Ber92}.
Alternative semiclassical quantization conditions based on a semiclassical
representation of the spectral staircase \cite{Aur92,Aur92b} and
derived from a quantum version of a classical Poincar\'e map \cite{Bog92}
are also restricted to bound and ergodic systems.

For systems with a symbolic dynamics the periodic orbit sum (\ref{g}) can
be reformulated as an infinite Euler product, which can be expanded
in terms of the cycle length of the symbolic code.
If the contributions of longer orbits are shadowed by the contributions
of short orbits the cycle expansion technique can remarkably improve the
convergence properties of the series and allows to extract the bound states
and resonances of bound and open systems, respectively
\cite{Cvi89,Art90,Eck93,Eck95}.
A combination of the cycle expansion technique with a functional equation 
for bound systems has been studied by Tanner et al.\ \cite{Tan91}.
However, the existence of a simple symbolic code is restricted to very few
systems, and cycle expansion techniques cannot be applied, e.g., to the general
class of systems with mixed regular-chaotic classical dynamics.

In this paper we present a general technique for the analytic continuation 
and the extraction of poles of a non-convergent series of the type of 
Eq.\ \ref{g}.
The method is based on {\em harmonic inversion} by filter-diagonalization.
The advantage of the method is that it does not depend on special 
properties of the system such as ergodicity or the existence
of a symbolic dynamics for periodic orbits.
It does not even require the knowledge of the mean staircase function, i.e.,
the Weyl term in dynamical systems.
The only assumption we have to make is that the analytic continuation of the
Dirichlet series $g(w)$ (Eq.\ \ref{g})
is a linear combination of poles $(w-w_k)^{-1}$, which is exactly the
functional form of, e.g., a quantum mechanical response function with
real and complex parameters $w_k$ representing the bound states and 
resonances of the system, respectively.
To demonstrate the general applicability and accuracy of our method we will 
apply it to two systems with completely different properties, first the zeros 
of the Riemann zeta function \cite{Edw74,Tit86}, as a mathematical model for 
a bound system, and second the three disk scattering system as a physical 
example.

As pointed out by Berry \cite{Ber86} the density of zeros of Riemann's zeta 
function can be written, in formal analogy with Gutzwiller's semiclassical
trace formula, as a non-convergent series, where the ``periodic orbits'' are
the prime numbers.
A special property of this system is the existence of a functional equation
which allows the calculation of Riemann zeros via the Riemann-Siegel formula 
\cite{Edw74,Tit86,Odl90}.
An analogous functional equation for quantum systems with an underlying 
chaotic (ergodic) classical dynamics has served as the basis for the 
development of a semiclassical quantization rule for bound ergodic systems
\cite{Ber86,Ber90,Kea92,Ber92}.
The Riemann zeta function has also served as a mathematical model to study
the statistical properties of level distributions \cite{Odl90,Boh84,Bog95}.
We will demonstrate that harmonic inversion can reveal the Riemann zeros
with extremely high accuracy and with just prime numbers as input data.
The most important advantage of our method is, however, its wide 
applicability, i.e., it can be generalized in a straightforward way to 
non-ergodic bound or open systems.

Our second example, the three disk scattering problem, is an open
and non-ergodic system.
Its classical dynamics is purely hyperbolic, and the periodic 
orbits can be classified by a complete binary symbolic code.
This system has served as a model for the development of cycle expansion
techniques \cite{Cvi89,Eck93,Eck95}.
When applying the harmonic inversion technique to the three disk scattering
system we will highlight the general applicability of our method by not 
having to make use of its symbolic dynamics in any way.

It is evident that methods invoking special properties of a given system
may be more efficient regarding, e.g., the number of periodic orbits
required for the calculation of a certain number of poles of the response
function $g(w)$ in that particular case.
It is not our purpose to compete with the efficiency of such methods.
Rather, the advantage of the harmonic inversion technique lies in its 
wide applicability, which allows the investigation also of systems not 
possessing special properties.
This is demonstrated in this paper by solving two completely different 
problems, viz.\ the zeros of the Riemann zeta function and the three disk 
scattering system, with one and the same method.

The paper is organized as follows.
In Section 2 we explain the general idea of the method by way of example of 
the Riemann zeros.
This is followed by the derivation of the harmonic inversion method in 
Section 3
and the presentation of numerical results for the Riemann zeros in Section 4.
The method is extended to the general case of periodic orbit quantization
in Section 5, and its usefulness and wide applicability is demonstrated for 
the three disk scattering system, as a physical example, in Section 6.

\section{The Riemann zeta function}
Our goal is to introduce our method for periodic orbit quantization
by harmonic inversion using, as an example, the well defined problem of 
calculating zeros of the Riemann zeta function.
There are essentially two advantages of studying the zeta function instead
of a ``real'' physical bound system.
First, the Riemann analogue of Gutzwiller's trace formula is exact,
as is the case for systems with constant negative curvature 
\cite{Gut90,Aur92b},
whereas the semiclassical trace formula for systems with plane geometry 
is correct only to first order in $\hbar$.
This allows a direct check on the precision of the method.
Second, no extensive periodic orbit search is necessary for the calculation
of Riemann zeros, as the only input data are just prime numbers.
It is not our intention to introduce yet another method for computing
Riemann zeros, which, as an objective in its own right, can be accomplished
more efficiently by specific procedures.
Rather, in our context the Riemann zeta function serves primarily as a 
mathematical model to illustrate the power of our technique when applied
to bound systems.

\subsection{General remarks}
Before discussing the harmonic inversion method we start with recapitulating 
a few brief remarks on Riemann's zeta function necessary for our purposes.
The hypothesis of Riemann is that all the non-trivial zeros of the analytic
continuation of the function
\begin{equation}
   \zeta(z)
 = \sum_{n=1}^\infty n^{-z}
 = \prod_p \left(1-p^{-z}\right)^{-1} \; , 
   \quad ({\rm Re}~ z>1,~ p: {\rm primes})
\label{zeta_def}
\end{equation}
have real part $1\over 2$, so that the values $w=w_k$, defined by
\begin{equation}
 \zeta\left({1\over 2}-iw_k\right) = 0,
\end{equation}
are all real or purely imaginary \cite{Edw74,Tit86}.
The Riemann staircase function for the zeros along the line
$z={1\over 2}-iw$, defined as
\begin{equation}
 N(w) = \sum_{k=1}^\infty \Theta(w-w_k),
\label{Riemann_staircase}
\end{equation}
i.e.\ the number $N(w)$ of 
zeros with $w_k<w$, can be split \cite{Ber86,Edw74,Tit86} into a smooth part, 
\begin{eqnarray}
   \overline N(w)
 &=& {1\over\pi}\arg\Gamma\left({1\over4}+{1\over2}iw\right)
   -{w\over 2\pi}\ln\pi + 1  \nonumber \\
 &=& {w\over 2\pi} \left(\ln \left\{{w\over 2\pi}\right\}-1\right)
   + {7\over 8} + {1\over 48\pi w} - {7\over 5760\pi w^3}+{\cal O}(w^{-5}) \; ,
\label{N_bar}
\end{eqnarray}
and a fluctuating part,
\begin{equation}
 N_{\rm osc}(w) = - {1\over \pi} \lim_{\eta\to 0} \, {\rm Im} \, 
  \ln \zeta\left({1\over 2}-i(w+i\eta)\right) \; .
\label{N_osc_def}
\end{equation}
Substituting the product formula (\ref{zeta_def}) (assuming that it can 
be used when ${\rm Re}~z={1\over 2}$) into (\ref{N_osc_def})
and expanding the logarithms yields
\begin{equation}
 N_{\rm osc}(w) = - {1\over \pi} \, {\rm Im} \, \sum_p \sum_{m=1}^\infty
 {1\over mp^{m/2}} \, e^{iwm\ln(p)} \; .
\end{equation}
Therefore the density of zeros along the line $z={1\over 2}-iw$ can
formally be written as
\begin{equation}
   \varrho_{\rm osc}(w)
 = \frac{dN_{\rm osc}}{dw}
 = -{1\over \pi} \, {\rm Im} \, g(w)
\label{rho_zeta}
\end{equation}
with the response function $g(w)$ given by the series 
\begin{equation}
g(w) = i\sum_p \sum_{m=1}^\infty \frac{\ln(p)}{p^{m/2}} \, e^{iwm\ln(p)} \ ,
\label{g_nc}
\end{equation}
which converges only for ${\rm Im}\, w > {1\over 2}$.

Obviously Eq.\ \ref{g_nc} is of the same type as the response function
(\ref{g}), with the entropy barrier $c={1\over2}$, i.e., Eq.\ \ref{g_nc} 
does not converge on the real axis, where the Riemann zeros are located.
The mathematical analogy between the above equation and Gutzwiller's
periodic orbit sum
\begin{equation}
 \varrho_{\rm osc}(E) \approx -{1\over \pi} \, {\rm Im} \, 
 \sum_{\rm po} {\cal A}_{\rm po} \, e^{iS_{\rm po}},
\label{rho_po}
\end{equation}
with ${\cal A}_{\rm po}$ the amplitudes and $S_{\rm po}$ the classical 
actions (including phase information) of the periodic orbit contributions,
was already pointed out by Berry \cite{Ber86,Ber90}.
For the Riemann zeta function the primitive periodic orbits have to be
identified with the primes $p$, and the integer $m$ formally counts the
``repetitions'' of orbits.
The ``amplitudes'' and ``actions'' are then given by 
\begin{eqnarray}
\label{Apm}
 {\cal A}_{pm}&=& i{\ln(p)\over p^{m/2}} \; , \\
\label{Spm}
 S_{pm}&=&mw\ln(p) \; .
\end{eqnarray}
Both equation (\ref{rho_zeta}) for the Riemann zeros and 
-- for most classically chaotic physical systems -- 
the periodic orbit sum (\ref{rho_po}) do not converge. 
In particular, zeros of the zeta function, or semiclassical 
eigenstates, cannot be obtained directly using these expressions.
The problem is to find the analytic continuation of these equations to the
region where the Riemann zeros or, for physical systems, the eigenenergies
and resonances, are located.
Eq.\ \ref{g_nc} is the starting point for our introduction and discussion
of the harmonic inversion technique for the example of the Riemann zeta
function.
The generalization of the method to periodic orbit quantization 
(Eq.\ \ref{rho_po}) in Section 5 will be straightforward.

Although Eq.\ \ref{g_nc} is the starting point for the harmonic inversion
method, for completeness we quote the Riemann-Siegel formula,
which is the most efficient approach to computing Riemann zeros.
For the Riemann zeta function it follows from a functional equation
\cite{Edw74} that the function
\begin{equation}
 Z(w) = \exp\left\{-i\left[\arg \Gamma\left({1\over 4}+{1\over 2}iw\right)
-{1\over 2}w\ln\pi\right]\right\}
        \zeta\left({1\over 2}-iw\right)
\end{equation}
is real, and even for real $w$.
The asymptotic representation of $Z(w)$ for large $w$,
\begin{eqnarray}
 Z(w) &=& -2\sum_{n=1}^{{\rm Int}[\sqrt{w/2\pi}]}
 \frac{\cos\{\pi\overline N(w)-w\ln n\}}{n^{1/2}} \nonumber \\
 &-& (-1)^{{\rm Int}[\sqrt{w/2\pi}]}\left({2\pi\over w}\right)^{1\over4} \;
   {\cos\left(2\pi\left(t^2-t-{1/16}\right)\right)
   \over\cos\left(2\pi t\right)} + \dots \; ,
\label{Riemann_Siegel}
\end{eqnarray}
with $t=\sqrt{w/2\pi}-{\rm Int}[\sqrt{w/2\pi}]$
is known as the Riemann-Siegel formula and has been employed (with several 
more correction terms) in effective methods for computing 
Riemann zeros \cite{Odl90}.
Note that the principal sum in (\ref{Riemann_Siegel}) has discontinuities
at integer positions of $\sqrt{w/2\pi}$, and therefore the Riemann zeros
obtained from the principal sum are correct only to about 1 to 15 percent
of the mean spacing between the zeros.
The higher order corrections to the principal Riemann-Siegel sum remove,
one by one, the discontinuities in successive derivatives of $Z(w)$ at the
truncation points and are thus essential to obtaining accurate numerical 
results.
An alternative method for improving the asymptotic representation of $Z(w)$
by smoothing the cut-offs with an error function and adding higher order 
correction terms is presented in \cite{Ber92}.

An analogue of the functional equation for bound and ergodic dynamical
systems has been used as the starting point to develop a ``rule for quantizing 
chaos'' via a ``Riemann-Siegel look-alike formula'' \cite{Ber90,Kea92,Ber92}.
This method is very efficient as it requires the least number of periodic
orbits, but unfortunately it is restricted to ergodic systems on principle
reasons, and cannot be generalized either to systems with regular or mixed 
classical dynamics or to open systems.

By contrast, the method of harmonic inversion does not have these restrictions.
We will demonstrate that Riemann zeros can be obtained directly from the 
"ingredients" of the non-convergent response function (\ref{g_nc}), i.e., 
the set of values $A_{pm}$ and $S_{pm}$, thus avoiding the use of the 
functional equation, the Riemann-Siegel formula, the mean staircase function 
(\ref{N_bar}), or any other special property of the zeta function.
The comparison of results in Section 4 will show that the accuracy of our 
method goes far beyond the Riemann-Siegel formula (\ref{Riemann_Siegel}) 
without higher order correction terms.
The main goal of this paper is to demonstrate that because of the formal 
equivalence between Eqs.\ (\ref{rho_zeta}) and (\ref{rho_po}) our method 
can then be applied to periodic orbit quantization of dynamical
systems \cite{Mai97b} without any modification.

\subsection{The ansatz for the Riemann zeros}
To find the analytic continuation of Eq.\ (\ref{g_nc}) in the region
${\rm Im}~ w<{1\over 2}$ we essentially wish to fit $g(w)$ to
its exact functional form, 
\begin{equation}
 g_{\rm ex}(w) = \sum_k {d_k \over w-w_k+i\epsilon} \; ,
\label{g_ex}
\end{equation}
arising from the definition of the Riemann staircase 
(\ref{Riemann_staircase}).
The ``multiplicities'' $d_k$ in Eq.\ \ref{g_ex} are formally fitting 
parameters, which here all should be equal to 1.

It is hard to directly adjust the non-convergent (on the real axis)
series $g(w)$ to the form of $g_{\rm ex}(w)$.
The first step towards the solution of the problem is to carry out the 
adjustment for the Fourier components of the response function,
\begin{equation}
  C(s) = {1\over 2\pi}\int_{-\infty}^{+\infty} g(w) e^{-isw}dw
= i\sum_p \sum_{m=1}^\infty \frac{\ln(p)}{p^{m/2}} \, \delta(s-m\ln(p)) \, ,
\label{C_nc}
\end{equation}
which after certain regularizations (see below) is a well-behaved function of 
$s$.
Due to the formal analogy with the results of periodic orbit theory 
(see Eqs.\ \ref{Apm} and \ref{Spm}), $C(s)$ can be interpreted as the 
recurrence function for the Riemann zeta function, with the recurrence 
positions $S_{pm}=m\ln(p)$ and recurrence strengths 
of periodic orbit returns $A_{pm}=i\ln(p)p^{-m/2}$.
The exact functional form which now should be used to adjust $C(s)$ is 
given by 
\begin{equation}
   C_{\rm ex}(s)
 = {1\over 2\pi}\int_{-\infty}^{+\infty} g_{\rm ex}(w) e^{-isw}dw
 = -i \sum_{k=1}^\infty d_k e^{-iw_ks} \; .
\label{C_ex}
\end{equation}
$C_{\rm ex}(s)$ is a superposition of sinusoidal functions with
frequencies
\footnote{It is convenient to use the word ``frequencies'' for $w_k$ 
referring to the sinusoidal form of $C(s)$. We will also use the word 
``poles'' in the context of the response function $g(w)$.}
$w_k$ given by the Riemann zeros and amplitudes $d_k=1$.

Fitting a signal $C(s)$ to the functional form of Eq.\ (\ref{C_ex}) with,
in general, both complex frequencies $w_k$ and amplitudes $d_k$ is known as
{\em harmonic inversion}, with a large variety of applications in various
fields \cite{Mar87,Wal95,Man97,Man97b,Mai97a}.
The harmonic inversion analysis is especially non-trivial if the number 
of frequencies in the signal $C(s)$ is large, e.g., more than a thousand. 
It is additionally complicated by the fact that the
conventional way to perform the spectral analysis by studying
the Fourier spectrum of $C(s)$ will bring us back to analyzing the 
non-convergent response function $g(w)$ defined in Eq.\ \ref{g_nc}.
Until recently the known techniques of spectral analysis \cite{Mar87} would 
not be applicable in the present case. 
It is the filter-diagonalization method \cite{Wal95,Man97,Man97b} which have
turned the harmonic inversion concept into a general and powerful 
computational tool.

The signal $C(s)$ as defined by Eq.\ \ref{C_nc} is not yet suitable for
the spectral analysis. 
The next step is to regularize $C(s)$ by convoluting it with a Gaussian 
function to obtain the smoothed signal,
\begin{eqnarray}
     C_\sigma(s)
 &=& {1\over \sqrt{2\pi}\sigma} \int_{-\infty}^{+\infty}
       C(s')e^{-(s-s')^2/2\sigma^2}ds' \nonumber \\
 &=& {i\over \sqrt{2\pi}\sigma} \sum_p \sum_{m=1}^\infty
       {\ln (p) \over p^{m/2}} \, e^{-(s-m\ln(p))^2 / 2\sigma^2}
\label{C_sigma}
\end{eqnarray}
that has to be adjusted to the functional form of the corresponding 
convolution of $C_{\rm ex}(s)$.
The latter is readily obtained by substituting $d_k$ in Eq.\ \ref{C_ex}
by the damped amplitudes,
\begin{equation}
 d_k \to d_k^{(\sigma)} = d_k \, e^{-w_k^2\sigma^2/2} \; .
\label{d_sigma}
\end{equation}
The regularization (\ref{C_sigma}) can also be interpreted as a cut of an 
infinite number of high frequencies in the signal which is of fundamental 
importance for numerically stable harmonic inversion.
Note that the convolution with the Gaussian function is no approximation,
and the obtained frequencies $w_k$ and amplitudes $d_k$ corrected by Eq.\
\ref{d_sigma} are still exact, i.e., do not depend on $\sigma$.
The convolution is therefore not related to the Gaussian smoothing devised
for Riemann zeros in \cite{Del66} and for quantum mechanics in \cite{Aur88},
which provides low resolution spectra only.

Before proceeding further we note that even though the derivation of 
Eq.\ \ref{C_sigma} assumed that the zeros $w_k$ are on the real axis,
the analytic properties of 
$C_\sigma(s)$ imply that its representation by Eq.\ \ref{C_sigma} 
includes not only the non-trivial real zeros, 
but also all the trivial ones, $w_k=-i(2k+{1\over 2})$, $k=1,2,\dots$,
which are purely imaginary.
The general harmonic inversion procedure described below does not
require the frequencies to be real. 
Both the real and imaginary zeros $w_k$ will be obtained as the eigenvalues 
of a non-Hermitian generalized eigenvalue problem.

\section{Filter-diagonalization method for harmonic inversion}
The harmonic inversion problem can be formulated as a non-linear fit 
(see, e.g., Ref.\ \cite{Mar87}) of the signal $C(s)$ defined on
an equidistant grid,
\begin{equation}
c_n\equiv C(n\tau)=\sum_k d_ke^{-in\tau w_k}\ ,\ \ 
n=0,1,2, \dots N,\label{eq:complexCt}
\end{equation}
with the set of generally complex variational parameters $\{w_k,d_k\}$.
(In this context the Discrete Fourier Transform scheme would 
correspond to a linear fit with $N$ amplitudes $d_k$ and 
fixed real frequencies $w_k=2\pi k/N\tau,\ k=1,2, \dots N$. The latter
implies the so called ``uncertainty principle'', i.e., the resolution, 
defined by the Fourier grid spacing, $\Delta w$, is inversely 
proportional to the length, $s_{\rm max}=N\tau$, of the signal C(s).)
The ``high resolution'' property associated with Eq.\ \ref{eq:complexCt} 
is due to the fact that there is no restriction for the closeness of the 
frequencies $w_k$ as they are variational parameters.
In Ref.\ \cite{Wal95} it was shown how this non-linear fitting problem can be
recast as a linear algebraic one using the filter-diagonalization procedure.
The essential idea is to associate the signal $c_n$ with an autocorrelation 
function of a suitable dynamical system, 
\begin{equation}
c_n=\left(\Phi_0,\hat U^n \Phi_0\right), \label{eq:Neuhc_n}
\end{equation}
where $(\ \cdot\ ,\ \cdot\ )$ defines a complex symmetric inner product 
(i.e., no complex conjugation).
The evolution operator can be defined implicitly by
\begin{equation}
   \hat U\equiv e^{-i\tau\hat\Omega}
 = \sum_{k=1}^K e^{-i\tau\omega_k} |\Upsilon_k)(\Upsilon_k| \; ,
\label{eq:U}
\end{equation}
where the set of eigenvectors $\{\Upsilon_k\}$ is associated with 
an arbitrary orthonormal basis set and the eigenvalues of $\hat U$ are 
$u_k\equiv e^{-i\tau\omega_k}$ (or equivalently the eigenvalues of 
$\hat \Omega$ are $\omega_k$).
Inserting Eq.\ \ref{eq:U} into Eq.\ \ref{eq:Neuhc_n} we obtain Eq.\ 
\ref{eq:complexCt} with 
\begin{equation}
d_k=(\Upsilon_k,\Phi_0)^2,\label{eq:dk_def}
\end{equation}
which also implicitly
defines the ``initial state'' $\Phi_0$.

This construction establishes an equivalence between
the problem of extracting spectral information from the signal with the one of
diagonalizing the evolution operator $\hat U=e^{-i\tau\hat \Omega}$ 
(or the Hamiltonian $\hat \Omega$) of the
fictitious underlying dynamical system. 
The filter-diagonalization method is then used for extracting the eigenvalues 
of the Hamiltonian $\hat \Omega$ in any chosen small energy window. 
Operationally this is done by solving a small generalized eigenvalue problem
whose eigenvalues yield the frequencies in a chosen window.
The knowledge of the operator $\hat \Omega$ itself is not required,
as for a properly chosen basis the matrix elements of $\hat \Omega$ can be 
expressed only in terms of $c_n$.
The advantage of the  filter-diagonalization procedure 
is its numerical stability with respect to both the length
and complexity 
(the number and density of  the contributing frequencies) of the signal.
Here we apply the method of Ref.\ \cite{Man97} which is an improvement
of the filter-diagonalization method of Ref.\ \cite{Wal95} in that it
allows to significantly reduce the required length of the signal by 
implementing a different Fourier-type basis
with an efficient rectangular filter. Such a basis
is defined by choosing a small set of values $\varphi_j$ in the frequency 
interval of interest,
$\tau\omega_{min}<\varphi_j<\tau\omega_{max},\ j=1,2,...,J$, and
the maximum order, $M$, of the Krylov vectors, 
$\Phi_n = e^{-in\tau\hat\Omega}\Phi_0$, used in the Fourier series,
\begin{equation} 
\Psi_j\equiv\Psi(\varphi_j)=\sum_{n=0}^M e^{in\varphi_j}\Phi_n
\equiv\sum_{n=0}^M e^{in(\varphi_j-\tau\hat\Omega)}\Phi_0.\label{eq:Psi_j}
\end{equation} 
It is convenient to introduce the notations, 
\begin{equation}
U^{(p)}_{jj'}\equiv  U^{(p)}(\varphi_j,\varphi_{j'})= 
\left(\Psi(\varphi_j),e^{-ip\tau\hat\Omega}\Psi(\varphi_{j'})\right), 
\label{eq:Up}
\end{equation}
for the matrix elements of the operator $e^{-ip\tau\hat\Omega}$,
and $\mbox{\bf U}^{(p)}$, for the corresponding small $J\times J$ complex 
symmetric matrix.
As such $\mbox{\bf U}^{(1)}$ denotes the
matrix representation of the operator $\hat U$ itself and $\mbox{\bf U}^{(0)}$,
the overlap matrix with elements $(\Psi(\varphi_j),\Psi(\varphi_{j'}))$,
which is required as the vectors $\Psi(\varphi_j)$ are not generally 
orthonormal.
Now using these definitions  we can set up a generalized eigenvalue problem,
\begin{equation}
\mbox{\bf U}^{(p)}\mbox{\bf B}_k=e^{-ip\tau\omega_k} 
\mbox{\bf U}^{(0)}\mbox{\bf B}_k, 
\label{eq:generalized}
\end{equation}
for the eigenvalues $e^{-ip\tau\omega_k}$ 
of the operator $e^{-ip\tau\hat\Omega}$. 
The column vectors $\mbox{\bf B}_k$ with elements $B_{jk}$ define 
the eigenvectors $\Upsilon_k$ in terms of the basis functions $\Psi_j$ as
\begin{equation}
\Upsilon_k = \sum_{j=1}^J B_{jk}\Psi_j, \label{eq:Bk}
\end{equation}
assuming that the $\Psi_j$'s form a locally complete basis.

The matrix elements (\ref{eq:Up}) can be expressed in terms of the signal 
$c_n$, the explicit knowledge of the auxiliary objects 
$\hat \Omega$, $\Upsilon_k$ or $\Phi_0$ is not needed.
Indeed, insertion of Eq.\ \ref{eq:Psi_j} into Eq.\ \ref{eq:Up},
use of the symmetry property,
$\left(\Psi,\hat U\Phi\right)=\left(\hat U\Psi,\Phi\right)$, and
the definition of $c_n$, Eq.\ \ref{eq:Neuhc_n}, gives after some arithmetics
\begin{eqnarray}
\label{eq:Upjj}
   U^{(p)} (\varphi,\varphi')
 &=& \big(e^{-i\varphi}-e^{-i\varphi'}\big)^{-1} 
     \Big[e^{-i\varphi}\sum_{n=0}^M e^{in\varphi'} c_{n+p}\\
 &-& e^{-i\varphi'}\sum_{n=0}^M e^{in\varphi}c_{n+p}
   -e^{iM\varphi}\sum_{n=M+1}^{2M}e^{i(n-M-1)\varphi'}c_{n+p}\nonumber\\
 &+& e^{iM\varphi'}\sum_{n=M+1}^{2M}e^{i(n-M-1)\varphi}c_{n+p}\Big],\ 
     \varphi\neq\varphi'\ ,\nonumber \\
   U^{(p)} (\varphi,\varphi)
 &=& \sum_{n=0}^{2M}(M-|M-n|+1)e^{in\varphi}c_{n+p}.\nonumber
\end{eqnarray}
(Note that the evaluation of {\bf U}$^{(p)}$ requires knowledge of $c_n$ for 
$n=p,p+1, \dots ,N=2M+p$.)

The solution of the generalized eigenvalue problem (\ref{eq:generalized})
is usually done by a singular value decomposition of the matrix 
{\bf U}$^{(0)}$. 
Each value of $p$ yields a set of frequencies $w_k$ and,
due to Eqs.\ \ref{eq:dk_def}, \ref{eq:Psi_j} and \ref{eq:Bk}, 
amplitudes, 
\begin{equation}
d_k=\left(\sum_{j=1}^J B_{jk}\sum_{n=0}^M c_ne^{in\varphi_j}\right)^2. 
\label{eq:dk}
\end{equation}
Note that Eq.\ \ref{eq:dk} is a functional of the half signal $c_n, 
n=1,2,\dots,M$.
Even though in all our applications Eq.\ \ref{eq:dk} yields very good results,
here we present an even better expression for the coefficients $d_k$ 
(see Ref.\ \cite{Man97b}), 
\begin{eqnarray}
d_k & = & \left[{1\over M+1}\sum_{j=1}^{J}B_{jk}
{\Big(}\Psi(\varphi_j),\Psi(w_k){\Big)}\right]^2\nonumber\\
& \equiv & \left[{1\over M+1}\sum_{j=1}^{J}B_{jk}U^{(0)}
  (\varphi_j,w_k)\right]^2
\label{eq:bestdk}
\end{eqnarray}
with $U^{(0)}(\varphi_j,w_k)$ defined by Eq.\ \ref{eq:Upjj}.
Eq.\ \ref{eq:bestdk} is a functional of the whole available signal $c_n, \ 
n=0,1,\dots,2M$.

The converged $w_k$ and $d_k$ should not depend on $p$. 
This condition allows us to identify spurious or non-converged frequencies 
by comparing the results with different values of $p$ 
(e.g., with $p=1$ and $p=2$).
We can define the simplest error estimate $\varepsilon$ as the difference 
between the frequencies $w_k$ obtained from diagonalizations with $p=1$ and 
$p=2$, i.e.\
\begin{equation}
 \varepsilon = | w_k^{(p=1)} - w_k^{(p=2)} | \; .
\label{eps_def}
\end{equation}

\section{Riemann zeros by harmonic inversion}
We have introduced in Section 2 the method for periodic orbit quantization
by harmonic inversion for the example of the Riemann zeta function because
this model allows a direct check of the precision of our method.
In this Section we present and discuss the numerical results obtained for 
the Riemann zeros.
We also discuss the amount of periodic orbit input data, viz.\ here 
the prime numbers, required to obtain converged semiclassical eigenenergies 
and Riemann zeros, respectively.

\subsection{Numerical results}
For a numerical demonstration we construct the signal $C_\sigma(s)$ using 
Eqs.\ \ref{C_nc} and \ref{C_sigma}
in the region $s<\ln(10^6)=13.82$ from the first 78498 prime numbers
and with a Gaussian smoothing width $\sigma=0.0003$.
Parts of the signal are presented in Fig.\ 1.
Up to $s\approx 8$ the Gaussian approximations to the $\delta$-functions
do essentially not overlap (see Fig.\ 1a) whereas for $s\gg 8$ the mean
spacing $\Delta s$ between successive $\delta$-functions becomes much less 
than the Gaussian width $\sigma=0.0003$ and the signal fluctuates around 
the mean $\overline C(s)=ie^{s/2}$ (see Fig.\ 1b).
From this signal we were able to calculate about 2600 Riemann zeros
to at least 12 digit precision.
For the small generalized eigenvalue problem (\ref{eq:generalized}) 
we used matrices with dimension $J<100$.
Some Riemann zeros $w_k$, the corresponding amplitudes $d_k$, and the 
estimated errors $\varepsilon$ (see Eq.\ (\ref{eps_def})) are given in 
Table 1.
Within the numerical error the Riemann zeros are real and the amplitudes
are consistent with $d_k=1$ for non-degenerate zeros.
To fully appreciate the accuracy of our harmonic inversion technique we note
that zeros obtained from the principal sum of the Riemann-Siegel formula 
(\ref{Riemann_Siegel}) deviate by about 1 to 15 percent of the mean spacing
from the exact zeros.
Including the first correction term in (\ref{Riemann_Siegel}) the 
approximations to the first five zeros read 
$w_1=14.137$, $w_2=21.024$, $w_3=25.018$, $w_4=30.428$, and $w_5=32.933$,
which still significantly deviates from the exact values (see Table 1).
Considering even higher order correction terms the results will certainly
converge to the exact zeros.
However, the generalization of such higher order corrections to ergodic
dynamical systems is a nontrivial task and requires, e.g., the knowledge
of the terms in the Weyl series, i.e., the mean staircase function after the 
constant \cite{Ber92,Kea94}.
The perfect agreement of our results for the $w_k$ with the exact Riemann 
zeros to full numerical precision is remarkable and clearly demonstrates 
that harmonic inversion by filter-diagonalization is a very powerful and 
accurate technique for the analytic continuation and the extraction of 
poles of a non-convergent series such as Eq.\ \ref{g}.

A few $w_k$ have been obtained (see Table 2) which are definitely 
not located on the real axis.
Except for the first at $w=i/2$ they can be identified with the trivial real 
zeros of the zeta function at $z=-2n$; $n=1,2,\dots$
In contrast to the nontrivial zeros with real $w_k$, the numerical accuracy 
for the trivial zeros decreases rapidly with increasing $n$.
The trivial zeros $w_n=-i(2n+{1\over2})$ are the analogue of resonances
in open physical systems with widths increasing with $n$.
The fact that the trivial Riemann zeros are obtained emphasizes the
general applicability of our method and demonstrates that periodic orbit 
quantization by harmonic inversion can be applied not only to closed but to 
open systems as well.
The decrease of the numerical accuracy for very broad resonances is a 
natural numerical consequence of the harmonic inversion procedure 
\cite{Man97,Man97b}.

The value $w=i/2$ in Table 2 is special because in this case the amplitude
is negative, i.e., $d_k=-1$.
Writing the zeta function in the form \cite{Tit86}
\begin{equation}
 \zeta({1\over 2}-iw) = C \prod_k (w-w_k)^{d_k} A(w,w_k)
\end{equation}
where $C$ is a constant and $A$ a regularizing function which ensures
convergence of the product, integer values $d_k$ are the multiplicities of 
{\em zeros}.
Therefore it is reasonable to relate negative integer values with the 
multiplicities of {\em poles}.
In fact, $\zeta(z)$ has a simple pole at $z={1\over 2}-iw=1$ consistent with
$w=i/2$ in Table 2.

\subsection{Required signal length}
We have calculated Riemann zeros by harmonic inversion of the signal
$C_\sigma(s)$ (Eq.\ \ref{C_sigma}) which uses prime numbers as input.
The question arises what are the requirements on the signal $C_\sigma(s)$,
in particular what is the required signal length.
In other words, how many Riemann zeros (or semiclassical eigenenergies)
can be converged for a given set of prime numbers (or periodic orbits,
respectively).
The answer can be directly obtained from the requirements on the harmonic
inversion technique.
In general, the required signal length $s_{\rm max}$ for harmonic inversion 
is related to the average density of frequencies $\overline\varrho(w)$ 
by \cite{Man97b}
\begin{equation}
 s_{\rm max} \approx 4\pi\overline\varrho(w) \; .
\label{s_max}
\end{equation}
From Eq.\ \ref{s_max} the required number of primes (or periodic orbits)
can be directly estimated as 
$\{\#~{\rm primes}~p~|~\ln p < s_{\rm max}\}$ or 
$\{\#~{\rm periodic~orbits}~|~s_{\rm po} < s_{\rm max}\}$.
For the special example of the Riemann zeta function the required number
of primes to have a given number of Riemann zeros converged can be estimated 
analytically.
With the average density of Riemann zeros derived from (\ref{N_bar}),
\begin{equation}
   \overline\varrho(w) = \frac{d\overline N}{dw}
 = {1\over 2\pi}\ln\left({w\over 2\pi}\right)
\end{equation}
we obtain
\begin{equation}
 s_{\rm max} = \ln(p_{\rm max}) = 2\ln\left({w\over 2\pi}\right)
 \Rightarrow p_{\rm max} = \left({w\over 2\pi}\right)^2 \; .
\end{equation}
The number of primes with $p<p_{\rm max}$ can be estimated from the
prime number theorem
\begin{equation}
   \pi(p_{\rm max}) \sim \frac{p_{\rm max}}{\ln(p_{\rm max})}
 = \frac{(w/2\pi)^2}{2\ln(w/2\pi)} \; .
\end{equation}
On the other hand the number of Riemann zeros as a function of $w$ is
given by Eq.\ (\ref{N_bar}).
The estimated number of Riemann zeros which can be obtained by harmonic 
inversion from a given set of primes is presented in Fig.\ 2.
For example, about 80 zeros $(w<200)$ can be extracted 
from the short signal $C_\sigma(s)$ with $s_{\rm max}=\ln(1000)=6.91$
(168 prime numbers) in agreement with the estimates given above.
Obviously, in the special case of the Riemann zeta function the efficiency 
of our method cannot compete with that of the Riemann-Siegel formula method
(\ref{Riemann_Siegel}) where the number of terms is given by
$n_{\rm max} = {\rm Int}\,[\sqrt{w/2\pi}]$ and, e.g., 5 terms in Eq.\
\ref{Riemann_Siegel} would be sufficient to calculate good approximations
to the Riemann zeros in the region $w<200$.
Our primary intention is to introduce harmonic inversion by way of example
of the zeros of the Riemann zeta function as a {\em general} tool for
periodic orbit quantization, and not to use it as an alternative method
for solving the problem of finding most efficiently zeros of the Riemann
zeta function.

A functional equation can only be invoked for the semiclassical quantization
of {\em bound} and {\em ergodic} systems.
In this case the required number of periodic orbits can be estimated from 
the condition $s_{\rm max} \approx \pi\overline\varrho(w)$
\cite{Ber90,Kea92,Ber92}, which differs by a factor of 4 from the required 
signal length (\ref{s_max}) for harmonic inversion.
Periodic orbit quantization by harmonic inversion will be of particular
advantage in situations where special properties such as a functional 
equation cannot be invoked, e.g., for bound systems with non-ergodic, i.e., 
regular or mixed classical dynamics, and for open (scattering) systems.

\subsection{A remark on the Riemann hypotheses}
We conclude this Section with a remark on the famous Riemann hypotheses
mentioned in Section 2.1.
Applying our method of harmonic inversion to the signal $C_\sigma(s)$
(Eq.\ \ref{C_sigma}) the Riemann hypotheses for the {\em zeros} of the 
zeta function, $\zeta(z={1\over 2}-iw_k)=0$, is directly related to an 
equivalent statement for the {\em eigenvalues} $e^{-i\tau w_k}$ of the 
operator $e^{-i\tau\hat\Omega}$, i.e., the generalized eigenvalue problem 
(\ref{eq:generalized}).
Speculations that the operator $\hat\Omega$ can be regarded as the 
Hamiltonian of a quantum mechanical system have been presented by Berry 
\cite{Ber86}.
Unfortunately, $\hat\Omega$ is not known as it is only defined implicitly 
by way of its matrix representation (\ref{eq:Upjj}), which is a linear
functional of the signal (\ref{C_sigma}).
However, the very fact that the Riemann zeros are obtained as eigenvalues
of some matrix with analytically known  coefficients is already intriguing.

\section{Periodic orbit quantization}
As mentioned in Section 2 the basic equation (\ref{g_nc}) used
for the calculation of Riemann zeros has the same mathematical
form as Gutzwiller's semiclassical trace formula.
Both series, Eq.\ \ref{g_nc} and the periodic orbit sum (\ref{rho_po}), 
suffer from similar convergence problems in that they are absolutely
convergent only in the complex half-plane outside the region where the
Riemann zeros, or quantum eigenvalues, respectively, are located.
As a consequence, in a direct summation of periodic orbit contributions 
smoothing techniques must be applied resulting in low resolution spectra
for the density of states \cite{Aur88,Win88}.
To extract individual eigenstates the semiclassical trace formula has to
be analytically continued to the region of the quantum poles.
Here dynamical zeta functions have turned out to be of particular interest.

For {\em bound} and {\em ergodic} systems one technique is to apply an 
approximate functional equation and generalize the Riemann-Siegel formula 
(\ref{Riemann_Siegel}) to dynamical zeta functions \cite{Ber90,Kea92,Ber92}.
The Riemann-Siegel look-alike formula has been applied, e.g., for the 
semiclassical quantization of the hyperbola billiard \cite{Kea94}.
For bound ergodic systems alternative semiclassical quantization conditions 
based on a semiclassical representation of the spectral staircase 
${\cal N}(E)=\sum_n\Theta(E-E_n)$ \cite{Aur92,Aur92b} and derived from a 
quantum version of a classical Poincar\'e map \cite{Bog92} have also been 
discussed.

These quantization techniques cannot be applied to {\em open} systems.
However, if a symbolic dynamics for the system exists, i.e., if the
periodic orbits can be classified with the help of a complete symbolic code,
the dynamical zeta function, given as an infinite Euler product over entries 
from classical periodic orbits can be expanded in terms of the cycle length 
of the orbits \cite{Cvi89,Art90}.
The {\em cycle expansion} series is rapidly convergent if the contributions 
of long orbits are approximately shadowed by contributions of short orbits.
The cycle expansion technique has been applied, e.g., to the three disk
scattering system \cite{Cvi89,Eck93,Eck95}, the three body Coulomb system 
\cite{Ezr91,Win92}, and to the hydrogen atom in a magnetic field \cite{Tan96}.
A combination of the cycle-expansion method with a functional equation has
been applied to bound systems in \cite{Tan91,Tan92}.
However, the existence of a complete symbolic dynamics is more the exception
than the rule, and the cycle expansion cannot be applied, in particular for 
systems with mixed regular-chaotic classical dynamics.

In this Section we apply the same technique that we used for the calculation 
of Riemann zeros, to the calculation of semiclassical eigenenergies 
and resonances of physical systems by harmonic inversion of Gutzwiller's 
periodic orbit sum for the propagator.
The method only requires the knowledge of all orbits up to a sufficiently 
long but finite period and does not rely on either an approximate semiclassical
functional equation, nor does it depend on the existence of a symbolic code 
for the orbits.
The method will therefore allow the investigation of a large variety of 
systems with an underlying chaotic, mixed, or even regular classical dynamics.
The derivation of an expression for the recurrence function to be harmonically
inverted is analogous to that in Section 2.2.

\subsection{Semiclassical density of states}
Following Gutzwiller \cite{Gut67,Gut90} the semiclassical response function
for chaotic systems is given by
\begin{equation}
 g^{\rm sc}(E) = g^{\rm sc}_0(E)
   + \sum_{\rm po} {\cal A}_{\rm po} e^{iS_{\rm po}} \; ,
\label{g_sc}
\end{equation}
where $g^{\rm sc}_0(E)$ is a smooth function and the $S_{\rm po}$ and 
${\cal A}_{\rm po}$ are the classical actions and weights (including phase 
information given by the Maslov index) of periodic orbit contributions.
Eq.\ (\ref{g_sc}) is also valid for integrable \cite{Ber76} and
near-integrable \cite{Tom95,Ulm96} systems but with different expressions 
for the amplitudes ${\cal A}_{\rm po}$.
It should also be possible to include complex ``ghost'' orbits 
\cite{Kus93,Mai97} and uniform semiclassical approximations
\cite{Alm87,Mai98} close to bifurcations of periodic orbits
in the semiclassical response function (\ref{g_sc}).
The eigenenergies and resonances are the poles of the response function
but, unfortunately, its semiclassical approximation (\ref{g_sc}) does
not converge in the region of the poles, whence the problem is the analytic 
continuation of $g^{\rm sc}(E)$ to this region.

In the following we make the (weak) assumption that the classical system has 
a scaling property, i.e., the shape of periodic orbits does not depend on 
the scaling parameter, $w$, and the classical action scales as 
\begin{equation}
 S_{\rm po} = ws_{\rm po} \; .
\label{S_po}
\end{equation}
Examples of scaling systems are billiards \cite{Cvi89,Hel84}, 
Hamiltonians with homogeneous potentials \cite{Mar89,Tom91}, 
Coulomb systems \cite{Win92}, or the hydrogen atom in external magnetic 
and electric fields \cite{Mai94,Tan96}.
Eq.\ \ref{S_po} can even be applied for non-scaling, e.g., molecular systems
if a generalized scaling parameter $w\equiv\hbar_{\rm eff}^{-1}$ is introduced
as a new dynamical variable \cite{Mai97c}.
Quantization yields bound states or resonances, $w_k$, for the scaling 
parameter.
In scaling systems the semiclassical response function $g^{\rm sc}(w)$ can be 
Fourier transformed easily to obtain the semiclassical trace of the propagator
\begin{equation}
   C^{\rm sc}(s) 
 = {1 \over 2\pi} \int_{-\infty}^{+\infty} g^{\rm sc}(w) e^{-isw} dw
 = \sum_{\rm po} {\cal A}_{\rm po} \delta\left(s-s_{\rm po}\right) \; .
\label{C_sc}
\end{equation}
The signal $C^{\rm sc}(s)$ has $\delta$-peaks at the positions of the 
classical periods (scaled actions) $s=s_{\rm po}$ of periodic orbits and 
with peak heights (recurrence strengths) ${\cal A}_{\rm po}$, i.e., 
$C^{\rm sc}(s)$ is Gutzwiller's periodic orbit recurrence function.
Consider now the quantum mechanical counterparts of $g^{\rm sc}(w)$ and
$C^{\rm sc}(w)$ taken as the sums over the poles $w_k$ of the Green's 
function,
\begin{equation}
 g^{\rm qm}(w) = \sum_k {d_k \over w-w_k+i\epsilon} \; ,
\label{g_qm}
\end{equation}
\begin{equation}
   C^{\rm qm}(s)
 = {1\over 2\pi} \int_{-\infty}^{+\infty} g^{\rm qm}(w) e^{-isw} dw
 = -i\sum_k d_k e^{-i w_k s} \; ,
\label{C_qm}
\end{equation}
with $d_k$ being the multiplicities of resonances, i.e., $d_k=1$ for 
non-degenerate states.
In analogy with the calculation of Riemann zeros from Eq.\ (\ref{C_sigma})
the frequencies, $w_k$, and amplitudes, $d_k$, can now be extracted by
harmonic inversion of the signal $C^{\rm sc}(s)$ after convoluting it
with a Gaussian function, i.e.,
\begin{equation}
   C_\sigma^{\rm sc}(s) 
 = {1\over \sqrt{2\pi}\sigma}
   \sum_{\rm po} {\cal A}_{\rm po} e^{(s-s_{\rm po})^2/2\sigma^2} \; .
\label{C_sc_sigma}
\end{equation}
By adjusting $C_\sigma^{\rm sc}(s)$ to the functional form of Eq.\ \ref{C_qm},
the frequencies, $w_k$, can be interpreted as the semiclassical approximation 
to the poles of the Green's function in (\ref{g_qm}).
Note that the harmonic inversion method described in Section 3 allows
studying signals with complex frequencies $w_k$ as well.
For open systems the complex frequencies can be interpreted as semiclassical 
resonances. 
Note also that the $w_k$ in general differ from the exact quantum eigenvalues
because Gutzwiller's trace formula (\ref{g_sc}) is an approximation, 
correct only to first order in $\hbar$.
Therefore the diagonalization of small matrices in (\ref{eq:generalized})
does not imply that the results of periodic orbit quantization are more
``quantum'' in any sense than those obtained, e.g., from a cycle expansion
\cite{Cvi89,Art90}.
The eigenvalues are solutions of non-linear equations and the diagonalization
is equivalent to the search of zeros of the dynamical zeta function in the
cycle expansion technique.
Numerical calculation of the zeros is also a non-linear problem and,
in contrast to the matrix diagonalization, might encounter a problem of
missing roots.

\subsection{Semiclassical matrix elements}
The procedure described above can be generalized in a straightforward manner
to the calculation of semiclassical diagonal matrix elements 
$\langle\psi_k|\hat A|\psi_k\rangle$ of a smooth Hermitian operator
$\hat A$.
In this case we start from the quantum mechanical trace formula \cite{Eck92}
\begin{equation}
 g_A^{\rm qm}(w) = {\rm tr} \, G^+ \hat A
 = \sum_k {\langle\psi_k|\hat A|\psi_k\rangle \over w-w_k+i\epsilon} \; ,
\label{gA_qm}
\end{equation}
which has the same functional form as (\ref{g_qm}), but with 
$d_k=\langle\psi_k|\hat A|\psi_k\rangle$ instead of $d_k=1$.
For the quantum response function $g_A^{\rm qm}(w)$ (Eq.\ \ref{gA_qm})
a semiclassical approximation has been derived in \cite{Eck92}, which
has the same form as Gutzwiller's trace formula (\ref{g_sc}) but
with amplitudes
\begin{equation}
 {\cal A}_{\rm po} = -i {A_p e^{-i{\pi\over 2}\mu_{\rm po}}
 \over \sqrt{|\det(M_{\rm po}-I)|}}
\label{A_po,A}
\end{equation}
where $M_{\rm po}$ is the monodromy matrix and $\mu_{\rm po}$ the Maslov
index of the periodic orbit, and
\begin{equation}
 A_p = \int_0^{S_p} A({\bf q}(s),{\bf p}(s)) ds
\label{A_cl}
\end{equation}
is the classical average of the observable $A$ over {\em one} period $S_p$ of
the {\em primitive} periodic orbit.
Note that ${\bf q}(s)$ and ${\bf p}(s)$ are functions of the classical action
instead of time for scaling systems \cite{Boo95}.
Gutzwiller's trace formula for the density of states is obtained with
$\hat A$ being the identity operator, i.e., $A_p=S_p$.
When the semiclassical signal $C^{\rm sc}(s)$ (Eq.\ \ref{C_sc}) with
amplitudes ${\cal A}_{\rm po}$ given by Eqs.\ \ref{A_po,A} and \ref{A_cl}
is analyzed with the method of harmonic inversion the frequencies and 
amplitudes obtained are the semiclassical approximations to the eigenvalues 
$w_k$ and matrix elements $d_k=\langle\psi_k|\hat A|\psi_k\rangle$,
respectively.

\section{The three disk scattering system}
Let us consider a billiard system consisting of three identical hard disks 
with unit radii, $R=1$, displaced from each other by the same distance $d$. 
This simple, albeit nontrivial, scattering system has served as a model for 
periodic orbit quantization in many investigations in recent years 
\cite{Gas89,Cvi89,Eck93,Eck95}.
After symmetry reduction the periodic orbits can be classified by a binary
symbolic code \cite{Cvi89}.
For $d>2.1$ there is a one-to-one identity between the periodic orbits and
the symbolic code, whereas for $d<2.1$ pruning of orbits sets in.
For $d=6$ semiclassical resonances were calculated by application
of the cycle expansion technique including all periodic orbits up to 
cycle length $n=13$ \cite{Eck95}. 
In order to demonstrate the usefulness of the harmonic inversion technique 
we first apply it to the case $R:d=1:6$ studied before.
Note that the ratio corresponds to the very favorable regime for the cycle 
expansion (see below).
In billiards the scaled action $s$ is given by the length $L$ of orbits 
$(s=L)$ and the quantized parameter is the absolute value of the wave
vector $k=|{\bf k}|=\sqrt{2mE}/\hbar$.
Fig.\ 3a shows the periodic orbit recurrence function, i.e., the trace of 
the semiclassical propagator $C^{\rm sc}(L)$.
The groups with oscillating sign belong to periodic orbits with adjacent
cycle lengths.
To obtain a smooth function on an equidistant grid, which is required for
the harmonic inversion method, the $\delta$-functions in (\ref{C_sc})
have been convoluted with a Gaussian function of width $\sigma=0.0015$.
As explained in Section 2 this does not change the underlying spectrum.
The results of the harmonic inversion analysis of this signal are
presented in Fig.\ 3b and Table 3.
The crosses in Fig.\ 3b represent semiclassical poles, for which the 
amplitudes $d_k$ are very close to $1$, mostly within one percent.
Because the amplitudes converge much slower than the frequencies these
resonance positions can be assumed to be very accurate within the semiclassical
approximation.
In fact, a perfect agreement to many significant figures is achieved for 
these poles with the results obtained by cycle expansion \cite{Eck95},
and this agreement confirms that the results in Fig.\ 3b and Table 3
are the true semiclassical resonances, i.e., deviations from the exact
quantum poles are solely due to the semiclassical approximation in
Gutzwiller's trace formula.
For some broad resonances marked by diamonds in Fig.\ 3b and Table 3
the $d_k$ deviate strongly from $1$, within 5 to maximal 50 percent.
It is not clear whether these strong deviations are due to numerical 
effects, such as convergence problems caused by too short a signal, or 
if they are a direct consequence of the semiclassical approximation.
Of course, in the exact expression (\ref{C_qm}) all multiplicities $d_k$ 
are 1, but there is no proof that this is still true within the semiclassical 
approximation.
However, for the lowest $k$ eigenvalues (see, e.g., the first three 
resonances in Table 3), where the agreement with the exact resonance energies 
is worst \cite{Eck95} $d_k=1$ still holds, indicating that there is no
$\hbar$-dependence for the multiplicities.

The cycle expansion technique is based on the idea that the contributions
of long periodic orbits are shadowed by those of short orbits.
For the three disk scattering system this is ideally fulfilled in the
limit of a large ratio $d/R\gg 2$.
This is no longer true for short distances between the disks and hence 
the convergence of the conventional cycle expansion becomes rather slow
\cite{Eck93}.
As a second example of periodic orbit quantization by harmonic inversion
we therefore study the three disk scattering system with a short distance
ratio $d/R=2.5$, and thus in a situation where the assumption of the
cycle expansion that contributions of long orbits are shadowed by short 
orbits is no longer a good approximation.
The results are presented in Fig.\ 4 and Table 4.
For large $L$ groups of orbits with the same cycle length of the symbolic 
code strongly overlap and cannot be recognized in Fig.\ 4a.
The signal is obtained from periodic orbits with cycle length $n\le 13$.
Note that only 356 periodic orbits with $L<7.5$ are included in the
signal (Fig.\ 4a) whereas the complete set of orbits with cycle length 
$n\le 13$ consists of 1377 orbits.
The resonances obtained by harmonic inversion of the semiclassical recurrence
function are in good agreement with results of the cycle expansion 
\cite{Wir97}.

We finally remark that when a symbolic dynamics exists and contributions 
of long orbits are shadowed by short orbits the cycle expansion techniques
are very efficient and, e.g., for the three disk scattering system with
$d\ge 6R$ a large number of resonances can be obtained from just 8 periodic
orbits with cycle length $n \le 4$.
Again, we do not aim at competing with the efficiency of the cycle expansion 
method in such ideal situations.
As already mentioned in Section 4 the advantage of periodic orbit 
quantization by harmonic inversion is its {\em general} applicability: 
it does not depend on the existence of a symbolic code and the shadowing of 
orbits.
We have extracted the semiclassical resonances of the three disk scattering 
problem directly from the periodic orbits with exactly the same method as 
the Riemann zeros in Section 4 from the primes as 'periodic orbits', for 
which no symbolic dynamics exists.

\section{Conclusion}
We have introduced harmonic inversion as a new and general tool for 
semiclassical periodic orbit quantization.
The method requires the complete set of periodic orbits up to a given maximum
period as input but does not depend on special properties of the orbits,
as, e.g., the existence of a symbolic code or a functional equation.
We have demonstrated the wide applicability of the method by applying it to
two systems with completely different properties, namely the zeros of the
Riemann zeta function and the three disk scattering problem.
Both systems have been treated before by efficient methods, which, however, 
are restricted to bound ergodic systems or systems with a complete 
symbolic dynamics.
The harmonic inversion technique allows to solve both problems with one and 
the same method.
Therefore the method can also serve as a tool for, e.g., the semiclassical 
quantization of systems with mixed regular-chaotic classical dynamics,
which still is a challenging and unsolved problem.
The signal $C^{\rm sc}(s)$ can be composed as the sum of a signal related
to the irregular part of the classical phase space with periodic orbit
amplitudes given by Gutzwiller's trace formula \cite{Gut90} and a signal
related to stable \cite{Ber76} or nearly integrable \cite{Tom95} torus 
structures.
It should also be possible to include, e.g., creeping orbits \cite{Wir92},
ghost orbit contributions \cite{Kus93,Mai97,Mai97a}, and higher order $\hbar$
corrections \cite{Gas93} into the signal $C^{\rm sc}(s)$, which can then
be inverted to reveal the semiclassical poles.
The method can even be used for a semiclassical periodic orbit quantization 
of systems with non-homogeneous potentials such as the potential surfaces of 
molecules when a generalized scaling technique \cite{Mai97c} is applied.

\ack
We are grateful to B.\ Eckhardt who kindly communicated to us his
periodic orbits for the three disk scattering system.
J.\ M.\ thanks the Alexander von Humboldt-Stiftung for a Feodor-Lynen
scholarship and H.\ Taylor and the University of Southern 
California for their kind hospitality and support.

\newpage
\section*{Figures and Tables}
%
\begin{figure}
\vspace{16.0cm}
\includegraphics{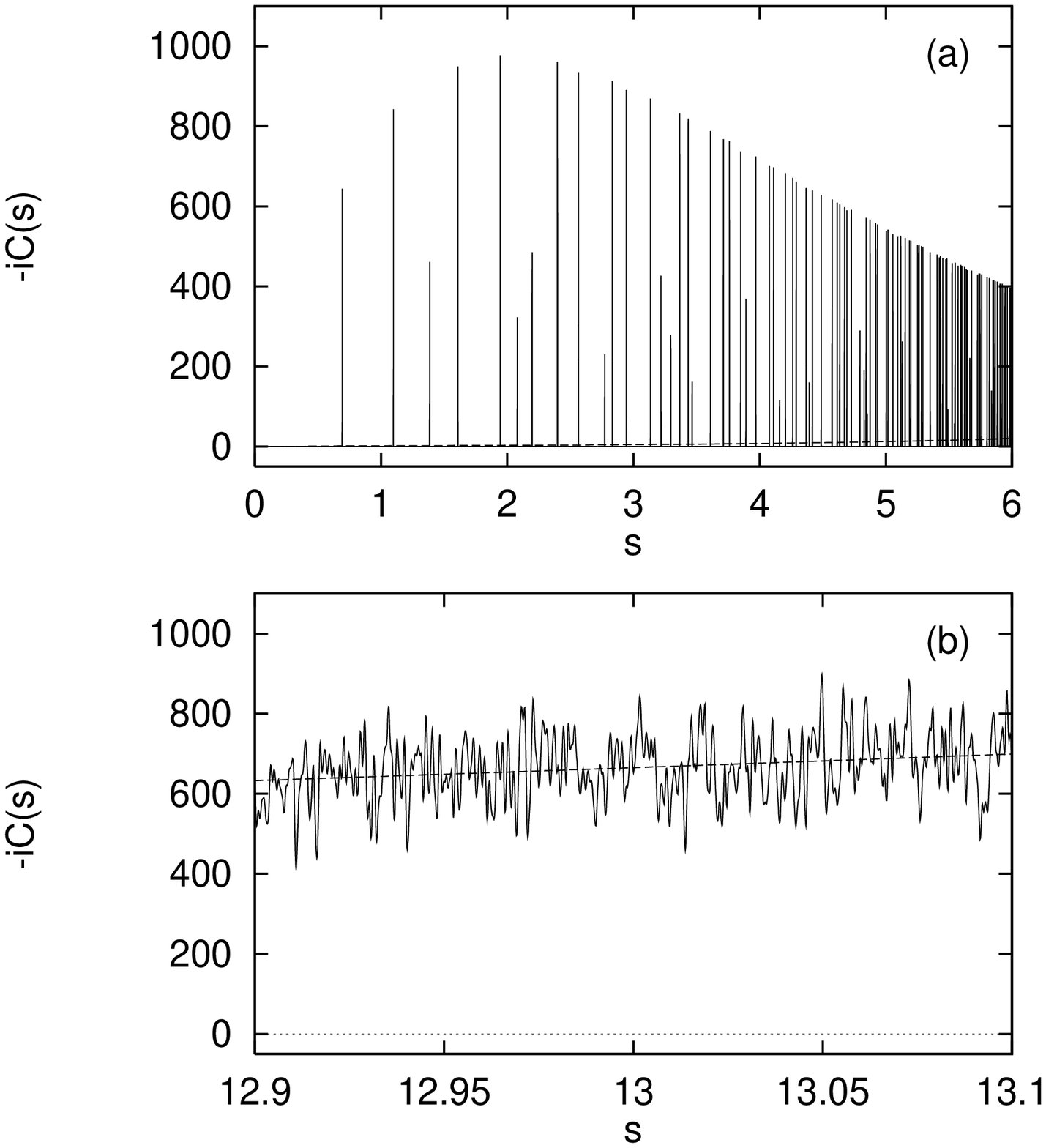}
\caption{\label{fig1} 
``Recurrence'' function $-iC_\sigma(s)$ for the Riemann zeros which has 
been analyzed by harmonic inversion. 
(a) Range $0\le s \le 6$, (b) short range around $s=13$.
The $\delta$-functions have been convoluted by a Gaussian with width 
$\sigma=0.0003$. Dashed line: Smooth background 
$\overline C(s)=ie^{s/2}$ resulting from the pole of the zeta function.
} 
\end{figure}
\begin{figure}
\vspace{9.5cm}
\includegraphics{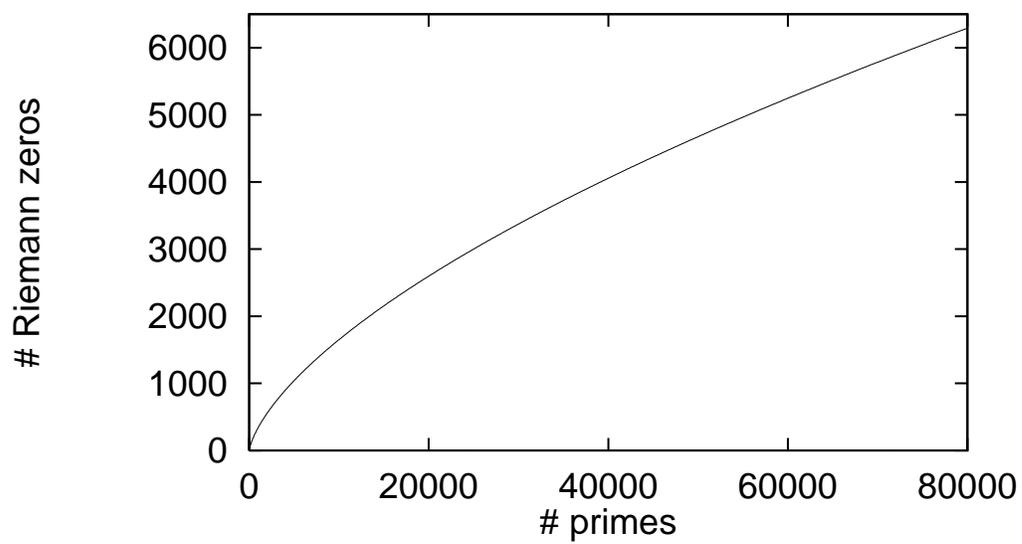}
\caption{\label{fig2} 
Estimated number of converged zeros of the Riemann zeta function, which 
can be obtained by harmonic inversion for given number of primes $p$.
}
\end{figure}
\begin{figure}
\vspace{19.5cm}
\includegraphics{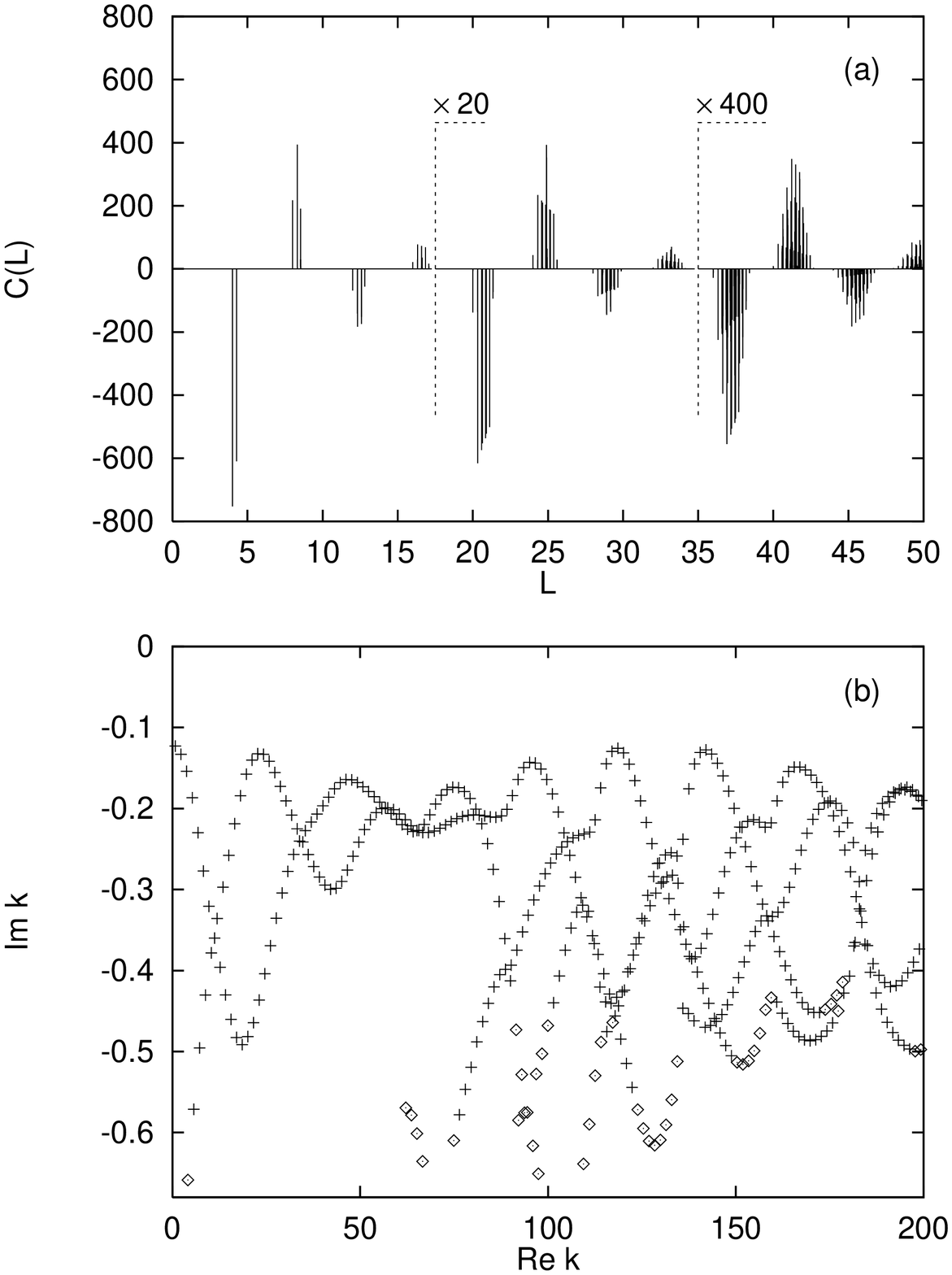}
\caption{\label{fig3} 
Three disk scattering system ($A_1$ subspace) with $R=1$, $d=6$.
(a) Periodic orbit recurrence function, $C(L)$.
The signal has been convoluted with a Gaussian of width $\sigma=0.0015$.
(b) Semiclassical resonances. The resonance positions marked by diamonds
might be less accurate (see text).
}
\end{figure}
\begin{figure}
\vspace{19.5cm}
\includegraphics{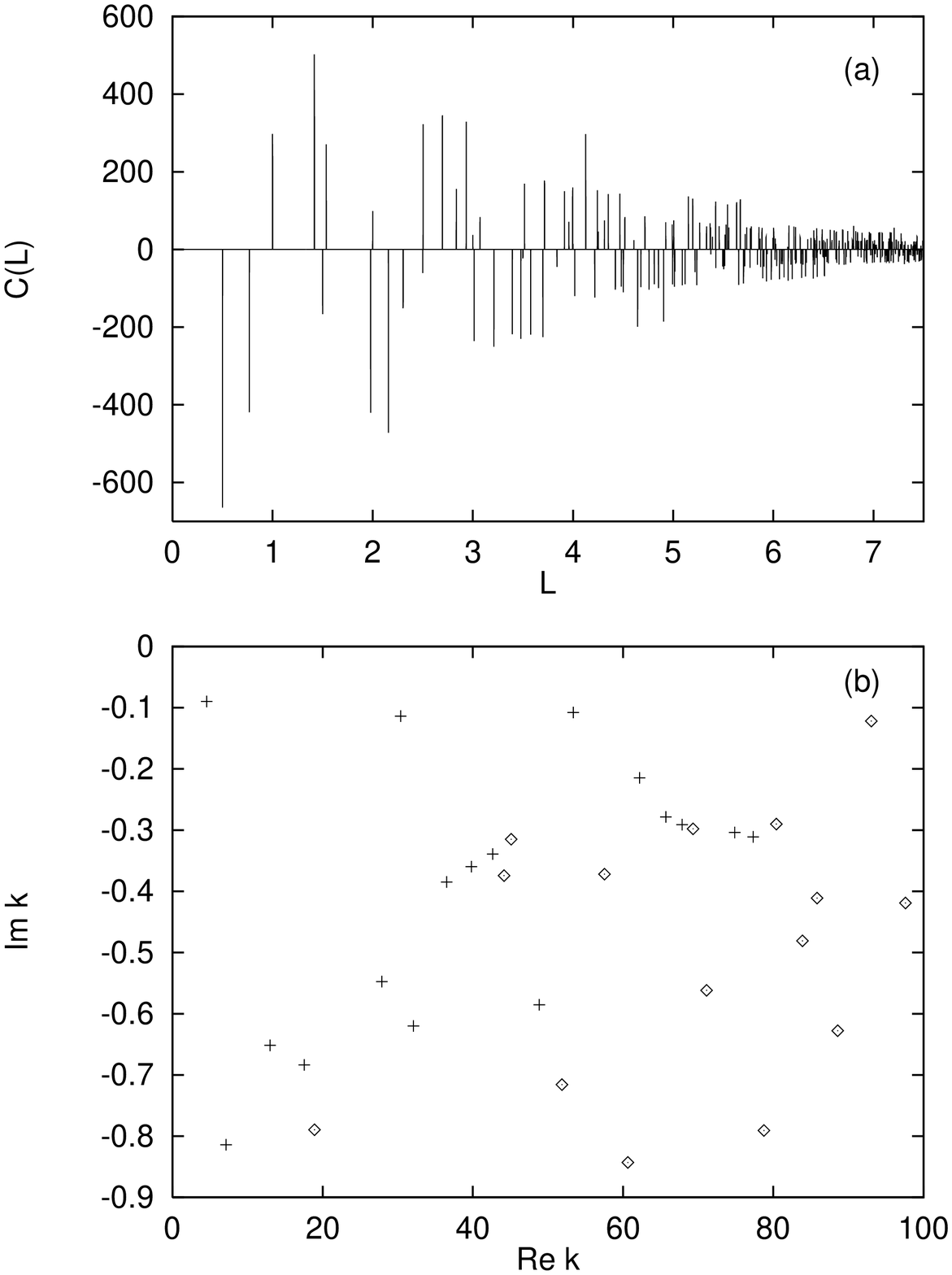}
\caption{\label{fig4} 
Three disk scattering system ($A_1$ subspace) with $R=1$, $d=2.5$.
(a) Periodic orbit recurrence function, $C(L)$.
The signal has been convoluted with a Gaussian of width $\sigma=0.0003$.
(b) Semiclassical resonances. The resonance positions marked by diamonds
might be less accurate (see text).
}
\end{figure}

\begin{table}[t]
\caption{\label{table1}
Non-trivial zeros $w_k$, multiplicities $d_k$, and error estimate 
$\varepsilon$ for the Riemann zeta function.}

\bigskip
\begin{center}
\begin{tabular}[t]{r||r|r||r|r||r}
  \multicolumn{1}{c||}{$k$} &
  \multicolumn{1}{c|}{${\rm Re}~w_k$} &
  \multicolumn{1}{c||}{${\rm Im}~w_k$} &
  \multicolumn{1}{c|}{${\rm Re}~d_k$} &
  \multicolumn{1}{c||}{${\rm Im}~d_k$} &
  \multicolumn{1}{c}{$\varepsilon$} \\ \hline
  1 &   14.13472514 &  4.05E-12 &     1.00000011 & -5.07E-08 &   3.90E-13 \\
  2 &   21.02203964 & -2.23E-12 &     1.00000014 &  1.62E-07 &   9.80E-13 \\
  3 &   25.01085758 &  1.66E-11 &     0.99999975 & -2.64E-07 &   5.20E-12 \\
  4 &   30.42487613 & -6.88E-11 &     0.99999981 & -1.65E-07 &   1.90E-12 \\
  5 &   32.93506159 &  7.62E-11 &     1.00000020 &  5.94E-08 &   7.10E-13 \\
  6 &   37.58617816 &  1.46E-10 &     1.00000034 &  5.13E-07 &   1.00E-12 \\
  7 &   40.91871901 & -3.14E-10 &     0.99999856 &  1.60E-06 &   4.90E-11 \\
  8 &   43.32707328 &  1.67E-11 &     1.00000008 &  3.29E-07 &   1.90E-12 \\
  9 &   48.00515088 &  4.35E-11 &     0.99999975 & -1.35E-07 &   1.40E-12 \\
 10 &   49.77383248 &  7.02E-11 &     1.00000254 & -4.59E-07 &   1.10E-10 \\
 11 &   52.97032148 &  1.92E-10 &     1.00000122 &  7.31E-07 &   6.00E-11 \\
 12 &   56.44624770 & -1.30E-10 &     0.99999993 &  4.51E-07 &   5.50E-12 \\
 13 &   59.34704400 &  5.40E-11 &     0.99999954 &  2.34E-06 &   2.30E-10 \\
 14 &   60.83177852 & -3.94E-10 &     1.00000014 &  1.11E-06 &   3.00E-11 \\
 15 &   65.11254406 & -4.98E-09 &     0.99998010 & -8.30E-06 &   2.70E-08 \\
$\cdots$ &  $\cdots$ &  $\cdots$ &  $\cdots$ &  $\cdots$ &  $\cdots$ \\
 2551 & 3083.36135798 &  8.43E-10 &     0.99999923 &  3.45E-07 &   1.50E-11 \\
 2552 & 3084.83845150 &  2.72E-09 &     1.00000057 & -2.86E-06 &   1.80E-10 \\
 2553 & 3085.37726898 & -1.37E-08 &     0.99999576 & -2.88E-06 &   5.50E-10 \\
 2554 & 3085.96552225 &  6.39E-09 &     0.99999667 &  1.50E-06 &   2.80E-10 \\
 2555 & 3087.01881535 &  3.46E-11 &     0.99999845 & -3.63E-07 &   5.20E-11 \\
 2556 & 3088.08343703 & -3.89E-10 &     0.99999931 & -8.44E-07 &   2.40E-11 \\
 2557 & 3089.22230894 & -3.31E-10 &     1.00000017 & -9.21E-07 &   1.80E-11 \\
 2558 & 3090.28219490 &  2.97E-10 &     1.00000069 & -7.17E-07 &   2.10E-11 \\
 2559 & 3091.15446969 &  1.10E-09 &     1.00000052 & -6.59E-07 &   1.50E-11 \\
 2560 & 3092.68766704 &  2.25E-09 &     1.00000033 &  1.45E-06 &   5.20E-11 \\
 2561 & 3093.18544571 & -2.33E-09 &     1.00000168 & -1.50E-07 &   6.40E-11 \\
 2562 & 3094.83306842 &  2.07E-08 &     0.99999647 &  2.63E-06 &   4.20E-10 \\
 2563 & 3095.13203122 & -1.79E-08 &     1.00000459 &  1.70E-06 &   5.20E-10 \\
 2564 & 3096.51548551 &  5.15E-09 &     0.99999868 &  2.74E-06 &   2.20E-10 \\
 2565 & 3097.34260655 &  7.75E-09 &     0.99999918 &  5.12E-06 &   6.00E-10 \\
 2566 & 3098.03835498 & -2.14E-08 &     1.00000296 &  4.34E-06 &   6.20E-10 \\
\end{tabular}
\end{center}
\end{table}

\begin{table}[t]
\caption{\label{table2}
Trivial zeros and pole of the Riemann zeta function.}

\bigskip
\begin{center}
\begin{tabular}[t]{r|r||r|r||r}
  \multicolumn{1}{c|}{${\rm Re}~w_k$} &
  \multicolumn{1}{c||}{${\rm Im}~w_k$} &
  \multicolumn{1}{c|}{${\rm Re}~d_k$} &
  \multicolumn{1}{c||}{${\rm Im}~d_k$} &
  \multicolumn{1}{c}{$\varepsilon$} \\ \hline
     0.00000000 &  0.50000000 &    -1.00000002 & -4.26E-08 &   1.80E-14 \\
    -0.00000060 & -2.49999941 &     0.99992487 & -3.66E-05 &   1.80E-07 \\
    -0.00129915 & -4.49987911 &     1.00069939 & -3.25E-03 &   4.40E-05 \\
    -0.09761173 & -6.53286064 &     1.07141445 & -1.49E-01 &   1.70E-03 \\
\end{tabular}
\end{center}
\end{table}

\begin{table}[t]
\caption{\label{table3}
Semiclassical resonances, multiplicities, and error estimates for the 
three disk scattering problem ($A_1$ subspace) with $R=1$, $d=6$. 
The marked resonances are plotted as diamonds in Fig.\ 3b.
}

\bigskip
\begin{center}
\begin{tabular}[t]{r|r||r|r||r}
  \multicolumn{1}{c|}{${\rm Re}~k$} &
  \multicolumn{1}{c||}{${\rm Im}~k$} &
  \multicolumn{1}{c|}{${\rm Re}~d$} &
  \multicolumn{1}{c||}{${\rm Im}~d$} &
  \multicolumn{1}{c}{$\varepsilon$} \\ \hline

    0.75831390 &   -0.12282220 &    0.99999998 &   -0.00000001 &   2.84E-12 \\
    2.27427857 &   -0.13305873 &    1.00000000 &    0.00000000 &   2.71E-14 \\
    3.78787678 &   -0.15412739 &    1.00000001 &    0.00000000 &   1.11E-13 \\
$\diamond\;$
    4.14568980 &   -0.65853972 &    0.94261284 &   -0.05782200 &   1.79E-06 \\
    5.29606778 &   -0.18678731 &    1.00000000 &    0.00000004 &   2.63E-12 \\
    5.68149760 &   -0.57137210 &    0.99512763 &   -0.01739098 &   5.34E-07 \\
    6.79363653 &   -0.22992212 &    0.99999994 &    0.00000018 &   1.54E-11 \\
    7.22405797 &   -0.49542427 &    1.00092001 &   -0.00461967 &   4.02E-07 \\
    8.27639062 &   -0.27708051 &    1.00000064 &   -0.00000007 &   3.47E-11 \\
    8.77921337 &   -0.43025611 &    0.99900081 &    0.00120544 &   1.00E-07 \\
    9.74763287 &   -0.32081704 &    0.99999986 &    0.00000049 &   6.32E-12 \\
   10.34422566 &   -0.37819884 &    1.00000189 &    0.00001109 &   3.89E-10 \\
   11.21347781 &   -0.35996394 &    1.00000180 &   -0.00000402 &   3.87E-10 \\
   11.91344955 &   -0.33573455 &    0.99999831 &   -0.00000066 &   2.50E-10 \\
   12.67753189 &   -0.39611536 &    0.99997860 &   -0.00000736 &   3.04E-09 \\
   13.48264892 &   -0.29694775 &    1.00000038 &   -0.00000054 &   6.90E-11 \\
   14.14241358 &   -0.43006040 &    1.00007802 &    0.00008149 &   2.54E-08 \\
$\cdots$ &  $\cdots$ &  $\cdots$ &  $\cdots$ &  $\cdots$ \\
  125.73060952 &   -0.33868744 &    1.00035374 &    0.00089953 &   1.76E-08 \\
  126.16812780 &   -0.21726568 &    0.99997532 &    0.00000523 &   6.34E-10 \\
  126.57000032 &   -0.30717994 &    0.99969830 &   -0.00028769 &   6.87E-09 \\
$\diamond\;$
  126.89863330 &   -0.61058335 &    1.24854908 &   -0.16290432 &   3.18E-06 \\
  127.21759681 &   -0.32010287 &    1.00042752 &    0.00045232 &   1.30E-08 \\
  127.68308651 &   -0.24341398 &    0.99993610 &   -0.00000236 &   2.13E-09 \\
  128.12116088 &   -0.28389637 &    1.00010175 &   -0.00022229 &   4.95E-09 \\
$\diamond\;$
  128.41137217 &   -0.61577414 &    1.32422538 &   -0.11565068 &   4.11E-06 \\
  128.70334065 &   -0.30442655 &    1.00039570 &    0.00011310 &   8.78E-09 \\
  129.19732946 &   -0.26788859 &    0.99987656 &   -0.00004493 &   5.25E-09 \\
  129.67319699 &   -0.26717842 &    1.00017817 &    0.00002664 &   4.62E-09 \\
$\diamond\;$
  129.92927207 &   -0.60918315 &    1.33322988 &   -0.02170382 &   4.76E-06 \\
  130.18796223 &   -0.29223540 &    1.00028313 &   -0.00012743 &   6.65E-09 \\
  130.71098079 &   -0.29045241 &    0.99983213 &   -0.00012609 &   8.19E-09 \\
  131.22717821 &   -0.25736473 &    1.00000071 &    0.00017324 &   5.54E-09 \\
$\diamond\;$
  131.44889208 &   -0.59054385 &    1.26610320 &    0.06973001 &   4.99E-06 \\
  131.67139581 &   -0.28429711 &    1.00010065 &   -0.00027624 &   6.60E-09 \\

\end{tabular}
\end{center}
\end{table}

\begin{table}[t]
\caption{\label{table6}
Semiclassical resonances, multiplicities, and error estimates for the 
three disk scattering problem ($A_1$ subspace) with $R=1$, $d=2.5$ 
obtained from the signal $C(L)$ with $L<7.5$ in Fig.\ 4a. 
The marked resonances are plotted as diamonds in Fig.\ 4b.
}

\bigskip
\begin{center}
\begin{tabular}[t]{r|r||r|r||r}
  \multicolumn{1}{c|}{${\rm Re}~k$} &
  \multicolumn{1}{c||}{${\rm Im}~k$} &
  \multicolumn{1}{c|}{${\rm Re}~d$} &
  \multicolumn{1}{c||}{${\rm Im}~d$} &
  \multicolumn{1}{c}{$\varepsilon$} \\ \hline

    4.58122247 &   -0.08999148 &    1.00000417 &    0.00120203 &   6.97E-08 \\
    7.14266960 &   -0.81391029 &    1.01834965 &   -0.00823113 &   3.51E-06 \\
   12.99951105 &   -0.65166427 &    1.00042048 &   -0.00056855 &   1.31E-06 \\
   17.56322689 &   -0.68356906 &    0.98459703 &   -0.03060711 &   2.28E-05 \\
$\diamond\;$
   18.91024231 &   -0.78956816 &    1.03497867 &   -0.07209360 &   6.09E-05 \\
   27.88792868 &   -0.54737225 &    1.02353431 &    0.00393194 &   2.30E-05 \\
   30.38871017 &   -0.11367391 &    1.00121948 &    0.00747117 &   2.68E-06 \\
   32.09837318 &   -0.62004177 &    0.99002266 &    0.01368255 &   3.50E-05 \\
   36.50721098 &   -0.38489303 &    1.00136214 &    0.01015558 &   9.58E-06 \\
   39.81154707 &   -0.35977801 &    1.00778064 &   -0.00596526 &   7.00E-05 \\
   42.65696984 &   -0.33910875 &    0.95316803 &    0.02942369 &   1.81E-05 \\
$\diamond\;$
   44.15561123 &   -0.37442367 &    0.70366781 &   -0.27927232 &   9.13E-05 \\
$\diamond\;$
   45.09670413 &   -0.31506305 &    0.77286261 &    0.10564855 &   2.54E-05 \\
   48.84367280 &   -0.58547564 &    0.97205165 &    0.01832879 &   3.83E-06 \\
$\diamond\;$
   51.85539738 &   -0.71582111 &    1.05998879 &   -0.22754826 &   7.06E-05 \\
   53.36884896 &   -0.10779998 &    1.03903158 &   -0.03195638 &   7.48E-06 \\
$\diamond\;$
   57.52623296 &   -0.37200326 &    0.54480246 &    0.04947934 &   6.95E-05 \\
$\diamond\;$
   60.63258604 &   -0.84290683 &    1.12993346 &    0.06170742 &   4.50E-05 \\
   62.20192292 &   -0.21464518 &    1.00384284 &    0.02041431 &   3.04E-06 \\
   65.68454001 &   -0.27861785 &    1.02584325 &    0.03420385 &   5.51E-06 \\
   67.86305728 &   -0.29098741 &    1.01748508 &   -0.01128120 &   6.19E-06 \\
$\diamond\;$
   69.32700248 &   -0.29789188 &    0.91331209 &   -0.06283737 &   5.73E-06 \\
$\diamond\;$
   71.11378807 &   -0.56166741 &    1.04431265 &    0.20817721 &   4.26E-06 \\
   74.85580547 &   -0.30392250 &    1.00774480 &    0.01774683 &   8.79E-07 \\
   77.31348462 &   -0.31110834 &    0.98293492 &   -0.00865456 &   5.06E-06 \\
$\diamond\;$
   78.74676605 &   -0.79088169 &    0.57558136 &   -0.30381968 &   1.39E-04 \\
$\diamond\;$
   80.39325912 &   -0.29001165 &    0.67726720 &   -0.02534165 &   3.99E-05 \\
$\diamond\;$
   83.89182348 &   -0.48077936 &    0.83497221 &    0.04875474 &   7.22E-05 \\
$\diamond\;$
   85.81836634 &   -0.41116853 &    0.97551535 &    0.10003934 &   1.05E-05 \\
$\diamond\;$
   88.57708414 &   -0.62777143 &    0.72880937 &    0.41114777 &   2.73E-05 \\
$\diamond\;$
   93.03487282 &   -0.12178427 &    0.98433404 &    0.07997697 &   9.37E-05 \\
$\diamond\;$
   97.58490354 &   -0.41923521 &    0.98663823 &    0.14408254 &   7.12E-05 \\

\end{tabular}
\end{center}
\end{table}

\end{document}